\documentstyle[12pt]{article}
\begin{document}
\textwidth 160mm
\textheight 240mm
\topmargin -20mm
\oddsidemargin 0pt
\evensidemargin 0pt
\newcommand{\beq}{\begin{equation}}
\newcommand{\eeq}{\end{equation}}

\begin{flushright}
ITEP-TH-49/00\\
hep-th/0010068\\

\end{flushright}

\begin{center}

{\bf RG flows on the phase spaces and the
$\tau$ functions for the generic Hamiltonian systems}

\vspace{1.5cm}
\centerline{A. S. Gorsky}
\vspace{0.2cm}

{\it Institute of Theoretical and  Experimental Physics, Moscow
117259, B. Cheryomushkinskaya 25}\\[0.2cm]

\end{center}
\vspace{2.0cm}

\begin{abstract}
We discuss the generic definition of the $\tau$ function for the
arbitrary Hamiltonian system. The different approaches concerning the
deformations of the curves and surfaces are compared. It is shown
that the Baker-Akhiezer function for the secondary integrable
system of the Toda lattice type can be identified with the coherent wave
function of the initial dynamical system. The $\tau$ function appears
to be related to the filling of the interior of the classical trajectory by
coherent states. Transition from  dispersionless to dispersionful
Toda lattice corresponds to the quantization of the initial
dynamical system.
\end{abstract}

\vspace{2.0cm}

1. Recently it was recognized that it is possible to attribute
the so called $\tau$ function to any analytic
curve in two dimensions \cite{mwz,wz}. This can be done
as follows. In the simplest case one  considers the evolution of the
curve when the area inside grows linearly
while the moments of the curve being fixed. In
more general situation all moments become
new variables ("times") too.   It
appears that in all cases the relevant object which encodes
the information on the properties of the deformation of the curve
is the quasiclassical $\tau$ function of the Toda lattice with the
additional constraint which selects the solution to the hierarchy.

On the other  hand it was shown in \cite{fgnr} that
it is possible to attribute the  prepotential
which is the quasiclassical version of the $\tau$
function to any holomorphic dynamical system. It plays
the role of the generating function for the
analogue of the S-duality transformation in the context of
Hamiltonian dynamics. Moreover it was argued
that generically there exists a pair of
dynamical systems such that their  coordinates
and actions get interchanged. Actually one can say
that along this viewpoint the
quasiclassical $\tau$ function can be attributed
to the Riemann surface which is the solution to the
equation of motion of the holomorphic system.

The third general setting involving prepotential
is the theory of the effective actions.
It appears that generically effective actions
manifest a lot of universality property
which is governed by the hidden integrability.
The examples are the identification
of the effective action of the N=2 SYM theory
with $\tau$ function of the quasiclassical hierarchy \cite{gkmmm}  and
similar integrable structures behind
d=2 theories which have been uncovered in  \cite{kawai, vafa, moore}.
Along this way  times in the hierarchy are identified with the
coupling constants in the corresponding theories. In more general
terms integrability amounts from the deformations of the
topological theories or topological sectors
in nontopological theories by the proper observables \cite{dubkr}.

The goal of this note is to attempt to glue these ideas altogether
and formulate the   place of $\tau$ function for the arbitrary
Hamiltonian system without any appealing to the  integrability.

2.
 Let us remind the key points from \cite{mwz}. One can consider
complex coordinates $\bar z, z$ and the curve determined by
the equation
\beq
\bar z=S(z)
\eeq
The Schwarz function  $S(z)$ is analytic in a domain including the curve.
One more ingredient to be defined is the map of the exterior of the curve
to the exterior of the unit disk
\beq
\omega (z)=\frac{z}{r} +\sum_{j} p_{j}z^{-j}
\eeq
where $\omega$ is defined on the unit circle. It is useful to introduce
the moments of the curve
\beq
t_n=\frac{1}{2\pi in} \oint z^{-n}S(z)dz
\eeq
\beq
v_n=\frac{1}{2\pi i} \oint z^nS(z)dz
\eeq
\beq
v_0=\oint log|z|dz
\eeq
which yield the following expansion for the Schwarz function
\beq
S(z)=\sum{k} t_k z^{k-1} +t_0 z^{-1} +\sum{k}v_k z^{-k-1}
\eeq
Let us define the generating function
\beq
S(z)=\partial_{z}\Omega(z)
\eeq
with the following expansion
\beq
\Omega(z)=\sum_{k=1} t_k z^{k} +t_0 logz -\sum_{k=1}v_k z^{-k}k^{-1} - 1/2 v_0
\eeq

One can easily prove the following relations
\beq
\partial_{t_0} \Omega(z)=\log \omega(z)
\eeq
\beq
\partial _{t_n} \Omega(z)=(z^n(\omega))_{+} +1/2 (z^n(\omega))_{0}
\eeq
\beq
\partial _{\bar{t_n}} \Omega(z)=(S^n(\omega))_{+} +1/2 (S^n(\omega))_{0}
\eeq
The symbol $(S(\omega))_{+}$ means the truncated Laurent series with only
positive powers of $\omega$  kept and the $(S(\omega))_{0}$
is the constant term in the series.
The differential $d\Omega$
\beq
d\Omega = Sdz + log\omega dt_0 +\sum (H_kdt_k -\bar {H_k} d \bar{t_k})
\eeq
provides the Hamiltonians and  $\Omega$ itself can be immediately
identified with the generating function for the canonical transformation
from the pair $(z,\bar {z})$ to the canonical pair $(t_0,log \omega)$.

The dynamical equations read
\beq
\partial _{t_n} S(z)= \partial _{z}H_n(z)
\eeq
\beq
\partial _{\bar{t_n}} S(z) = \partial _{z}\bar {H_n}(z)
\eeq
and the consistency of (13) (14) yields the zero-curvature condition
which amounts to the equations of the dispersionless Toda lattice
hierarchy. The first equation of the hierarchy reads as follows
\beq
\partial^{2}_{t_1\bar {t_1}}\phi = \partial_{t_0}e^{\partial_{t_0}\phi}
\eeq
where $ \partial_{t_0}\phi =2logr$.
The Lax operator L coincides with $z(\omega)$   and its eigenfunction -
Baker-Akhiezer function looks as follows
$\Psi=e^{\frac{\Omega}{h}}$. Hamiltonians
are expressed in terms of the Lax operator
\beq
H_k=(L^k)_{+} +1/2(L^k)_{0}
\eeq

3. In this section we shall interpret the evolution above
in terms of dynamics on the phase space assuming
that  $\bar z, z$  pair yields the phase space of some
dynamical system. The curve itself corresponds to the
energy level of this dynamical system. With this setup
it is clear that Poisson bracket between $\bar z$ and $z$
is fixed by the standard symplectic form . The next step
is the identification of $logw$ as angle variable hence
the identification of the area inside the curve $t_0$ as the
action variable is evident.

The holomorphic variables on the phase space are usually
considered as the operators in the Fock space so
we identify $z=b;\bar{z}=b^{+}$. Typically the creation-
annihilation operators are expressed it terms of the $(p,q)$ variables
however for our needs we assume the action-angle representation.
The variable $z=b$ can be expressed in terms of action-angle variables
in the particular dynamical system
and just this expression provides the Toda lattice Lax operator
\beq
z(logw,t_0)=L(logw,t_0)
\eeq
Baker-Akhiezer (BA) function is solution to the equation
\beq
L\Psi(z,t_0)=z\Psi
\eeq
and the complex $z$ plane is identified with the surface where the spectral
parameter of the Toda system lives on.

Now we turn to clarification of the meaning of the BA function
in the generic dynamical system. The answer follows from the
equation above; BA function is nothing but the coherent wave function
in the action representation. Indeed the coherent wave function
is the eigenfunction of the creation operator
\beq
\hat b \Psi=b\Psi
\eeq

The main properties  of the coherent states can be summarized
as follows \cite{perelomov}.
The coherent state can be represented in the form
\beq
|\alpha>= D(\alpha)|0>
\eeq
where $|0>$ is the normalized vacuum vector annihilated by the operator b
and the $D(\alpha)=exp(\alpha b^{+} -\bar{\alpha}b)$
yields the action of the Heisenberg-Weyl group. The operators $D(\alpha)$
obey the following relation
\beq
D(\alpha)D(\beta)=e^{2iIm(\alpha \bar{\beta})}D(\beta)D(\alpha)
\eeq
The coherent states $|\alpha>$ and $|\beta>$ are not orthogonal moreover
the system of the coherent states $[|\alpha>]$ is overcompleted
generically.  To get the complete system one can proceed as follows.
The lattice on the complex plane $\alpha_{nm}=n\omega_1 +m\omega_2 $
can be introduced with the area of the elementary cell equals
to the elementary Planck cell $\pi$.
In this case the system of the coherent states $|\alpha_{nm}>$
can be expressed in terms of the standard theta-functions on
the torus and turns out to be complete.

Note that the system of the coherent states can be defined
for the generic group as follows
\beq
|\psi_{g}>=T(g)|\psi_{0}>
\eeq
If H is the stabilizer of the vector $|\psi_{0}>$  the coherent state is
defined by the point x=x(g)  from
the homogeneous space G/H corresponding to the element g.
Moreover the state corresponding to $|x>$ can be identified with
the one dimensional projector $P_{x}=|x><x|$.

Now we are ready to recognize the meaning of the generating
function $\Omega$. From the equations above it becomes clear that
it is the generating function for the canonical transformations
from the $b,b^{+}$ representation to the angle-action variables.
Let us also remark that the role of the Orlov-Shulman operator
becomes transparent. This operator looks like $b^{+}b$ and
its eigenvalue counts the "number of particles".

Having identified the BA function for the generic system let us
consider the role of the $\tau$ function in the generic case. To this
aim it is convenient to use the following formulae for the
$\tau$ function
\beq
\tau(t,W)=<t,\bar{t}|W>
\eeq
where the bra vector depends on times while the ket vector
is fixed by the so called point of Grassmanian. One more useful
representation is provided by the fermionic language
\beq
\tau(t,W)=\frac{<N|\Psi(z_1)...\Psi(z_N)|W>}{\Delta(z)}
\eeq
where $\Delta(z)$ is Vandermonde determinant.

Let us briefly comment on the definition of the point
of the Grassmanian W. Generally speaking the Grassmanian
itself can be considered as a collection of all
fermionic Bogolyubov transforms of the vacuum state.
Hence we can say that W belongs to the Grassmanian
if it is annihilated by some linear combination
of the fermionic creation and annihilation operators.
Equivalently one could consider the following definition
\beq
|W>=S|0> , S=exp\sum_{nm}A_{nm}\bar{\psi}_{-n-1/2} \psi_{-m-1/2}
\eeq
that is S can be considered as the element of $GL(\infty,C)$.

The consideration above suggests the following picture behind the
definition of the $\tau$ function. The peculiar classical trajectory
of the dynamical system  yields the curve on the phase space. Then
the domain inside the trajectory is filled by the coherent states
for this particular system. Since the coherent state occupies the
minimal cell of the phase space the number of the coherent states
packed inside the domain is finite and equals N. Since there is only one
coherent state per cell for the complete set it actually behaves
like a fermion implying a kind of the fermionic representation.

Therefore we can develop the second dynamical system of the Toda
lattice type based on the generic dynamical system. The number
of the independent time variables in the Toda system  amounts from
the independent parameters in the potential in the initial
system plus additional time attributed to the action variable.
The situation is similar to the consideration of the
pair of the dual dynamical systems in \cite{fgnr}. The subtle
point in that paper is the choice of the Hamiltonians
for the dual systems. It seems that consideration in this note
provides the unified viewpoint on this issue. Indeed
formulae above  suggest the set of commuting Hamiltonians for the
dual system.  Let us emphasize that the choice of the particular
initial dynamical system amounts to the choice of the
particular solution to the Toda lattice hierarchy.

It is important that the energy level of the dynamical system
doesn't develop the discontinuity in some range of the
energies. This happens only if the system approaches
separatrices generically existing
on the phase space.

Consider the example of  the oscillator. Corresponding
phase trajectories
\beq
E=p^2 + q^2
\eeq
are circles on the phase space (ellipses in the generic case) and the
natural complex variables are
\beq
z=p+iq
\eeq
If we  calculate the expression for $v_0$
the following prepotential for the oscillator emerges
\beq
{\cal F}= 1/2 I^2 logI - 3/4 I^2
\eeq
where I is the action variable. To get the first
dispersionless Toda equation one has to consider
the phase trajectories with the shifted center. The position
of the center is parameterized by times $t_1,\bar{t_1}$. If the
Toda equation with three nonvanishing times is solved the prepotential
above can be reproduced. Let us also note that the
dual Hamiltonian in the sense of \cite{fgnr} is
\beq
H=b=\sqrt{t_0}e^{i\partial_{t_0}}
\eeq
The wave function of the dual system coincides with
the coherent wave function in the action representation.

The prepotential for the complex system is
defined by its spectral curve
\beq
a_D=\frac{\partial  {\cal F}}{\partial a}
\eeq
where $a_D$ and $a$  are the integrals of the
meromorphic differential over A and B cycles. The variable $a$
is just the action variable.  The S duality transformation
$a \rightarrow a_D$ actually maps the region of small energies
to the region of the large energies. Prepotential
obeys
the Matone  type \cite{matone}
relation which has the meaning of the Ward identity
\beq
\frac{\partial \cal F}{log \Lambda}=const H
\eeq
where H is the Hamiltonian of the complex system and $\Lambda$
is the scale factor.
The simple calculation shows that such relation doesn't`t hold
in the real case.

Let us remind how the prepotential provides the
S-duality transformation for the complex dynamical system \cite{fgnr}.
The action variables in dynamical system
are the integrals of meromorphic differential $\lambda$ over
the $A$-cycles on the spectral curve.
The reason for the $B$-cycles to be
discarded is simply the fact that the $B$-periods of $\lambda$
are not independent of the $A$-periods. On the other
hand, one can choose as the independent periods the
integrals of $\lambda$ over any lagrangian
subspace in $H_{1}({\bf T}_{b}; Z)$.

This leads to the following structure of the action variables
in the holomorphic setting. Locally over a disc in $B$ one
chooses a basis in $H_{1}$ of the fiber together with the
set of $A$-cycles. This choice may differ over another
disc. Over the intersection of these discs one has
a $Sp(2m, Z)$ transformation relating the bases
which has a natural interpretation as a S-duality
transformation. Altogether
they form an $Sp(2m, Z)$ bundle. It can be easily shown
the two form:
\beq
dI^{i} \wedge dI^{D}_{i}
\eeq
vanishes. Therefore one can always locally  find a function
$\cal{F}$ , such that:
\beq
I^{D}_{i} = {{\partial \cal{F}}\over{\partial I^{i}}}
\eeq
The angle variables are uniquely reconstructed once the action
variables are known. Since the real system to some
extend can be considered as a real section of the dynamical system
one could expect the analogue of the S-duality for the
real system too.

An important outcome from our identification is the
natural recipe for the transition from the
dispersionless to dispersionful Toda lattices.
Indeed at the  first step of quantization
the system one imposes the Bohr-Sommerfeld
quantization condition on the action variable. This is
in agreement with the discretization of $t_0$ variable
in Toda lattice. The Planck constant in the initial
system plays the transparent role in the Toda lattice.

4.
Let us argue now that there is clear correspondence
between the integrable dynamics behind the deformation
of the curves which are the phase space trajectories
for the real dynamical systems and  the deformation
of the surfaces which represent the trajectories of the
complex dynamical systems. Remind that
to formulate the integrable dynamics behind
the deformation of the curves one introduces
the following main objects; generating function $\Omega$, $\tau$
function depending on time variables $t_k$ which are the moments
of the  exterior of the classical trajectory as well as the
dual variables $v_k$ representing the moments of interior.

It is not a difficult task to recognize their counterparts
in the case of the surfaces. First let us represent the time
variables in the following form
\beq
v_k=\oint z^k d\Omega(z)
\eeq
\beq
t_k=\oint z^{-k} d\Omega(z)
\eeq
and compare it with the definition of times in the case of surfaces
\beq
T_n=res_{\xi=0}{\xi^n}dS_{SW}
\eeq
\beq
{{\partial \cal{F}}\over{\partial T^{n}}}= \frac{1}{2\pi in} res_{0}
\xi^{-n} dS_{SW}
\eeq
where $S_{SW}$ in the context of the N=2
SYM theories is the so-called Seiberg-Witten
meromorphic differential whose derivatives with respect of the moduli of the
curve are the holomorphic differentials.
We see that the variables in cases of curves and surfaces are differed
just by the Legendre transform.
The variable $t_0$
is nothing but the variable
\beq
a=\oint_{A}dS
\eeq
where the integration is over the A cycle on the spectral surface is the
complex dynamical system with only one degree of freedom is considered.
The definition of conjugated variables $v_n$ is  more delicate
issue since naively there are no the direct analogue of the
integrals over the B cycles
\beq
a_{D}=\oint_{B}dS
\eeq
in the real case. However it can be
seen that the analogous variable for  the one degree
of freedom can be identified
with variable $v_0$ from
\cite{mwz,wz}. If there are several degrees of freedom the
corresponding variables
$a_i$ are defined as the integrals over the complete set of A cycles.

It is interesting to recognize the real analogue of the modulus
of the spectral curve  in the complex case
\beq
\tau=\frac{da_{D}}{da}
\eeq
which plays the role of the effective coupling constant
in the field theory context. In the problem under consideration the
corresponding variable is
\beq
logr=\frac{dv_0}{dt_0}
\eeq
and also exhibits the  behaviour similar to
the perturbative regime in the field theory.
In what follows we shall see more indications that
dynamics of the curve has a pertubative regime as
a field theory counterpart.

In the context of the Seiberg-Witten solution to the N=2 SUSY YM
theory  the Whitham dynamics has a transparent renormalization
group meaning \cite{gmmm}. Indeed the derivative of the prepotential
with respect to the first time in SUSY YM context looks as follows
\beq
\frac {\partial \cal{F}}{\partial T_1}=
a\frac {\partial \cal{F}}{\partial a} -2{\cal{F}}=2H(a)
\eeq
where the term at the r.h.s. coincides with the Hamiltonian
for the dynamical system providing the corresponding spectral
curve. The dependence $H(a)$ on the action variable is the
intrinsic characteristics of the dynamical system, moreover
this term has an anomalous nature from the field theory viewpoint.
The analogous equation has been also found for the real case \cite{mwz}
where the observed quadratic dependence on the action
in the anomaly term corresponds to the
perturbative limit in the field theory framework.

Given such RG interpretation in the complex case one could look
for the reasonable RG structure behind the real dynamical system.
The following  starting point can be chosen. Assume that the
RG scale is fixed by the energy corresponding to the classical trajectory
or by the variable $t_0=\mu$.
Then the phase space is naturally divided into the "IR theory"
inside the trajectory and the "UV theory" outside. Therefore
a kind of the RG problem can be formulated. To this aim
introduce the classical expectation value of the observable
in the theory
\beq
<O>=\int dpdq O(p,q) \rho(p,q)
\eeq
where $\rho(p,q)$ is the phase space density (Wigner function). The
vacuum expectation value at the scale $\mu$  is naturally
introduced if the integration over the phase space up to the
energy $\mu$ is performed.

Now we can formulate the following observation - correlators
in UV theory in the outer region amount to the
coupling constants in the inner region and vice versa
\beq
\tau(t_0,t_k)=<exp\sum t_k O_k>
\eeq
where the  matrix element is taken in the IR theory.
The key point
is that  usually effective action
$S(\mu)$  identified with the prepotential
is defined via integrating
out all  modes up to the scale $\mu$. From \cite{mwz}
it becomes clear that is has to be done in a clever way;
one has to integrate out the part of the phase space
outside of the phase trajectory corresponding to the
energy E. Then the Riemann-Hilbert problem is defined on the
RG scale $\mu$ - one has to divide the whole theory into UV and IR
parts consistently with the RG flows. Remark that
interpretation of the area inside the curve
as the RG parameter agrees with the interpretation of the
size of the matrix in the matrix model as RG scale \cite{brezin}.
Let us emphasize that the picture emerged is formally
identical to the consideration in \cite{gmmm} where $\tau$
function approach amounts to the explicit calculation of the
RG dependence of the correlators. Along this viewpoint
one could expect that the similar correlators which now
carry the information about the spectrum of the system
could be found. Therefore instead of the determination
of the RG invariant characteristics of the field
theories we expect the RG like evaluation of the spectral invariants
which would distinguish different behaviour
of the Hamiltonian systems.

Let us emphasize that the Toda lattice equations are written
for the variable $b(\omega)$ which yields the dependence
of the creation operator on the angle variable. If only
zero time is involved then the evolution
doesn`t change the system however with all times switched on
the dynamics proceeds in the space of the dynamical systems
or can be interpreted as the RG flows.
The integrals of motion in the Toda system
are the "classical correlators" in the IR theory.

5.
It was shown in \cite{mwz} that the $\tau$ function of the curve allows
two additional complementary interpretations. First,  it has the
interpretation of the energy of the fermions with the unit charge
inside the curve with the Coulomb interaction taken into account.
Secondly it can be considered as the partition function of the
normal matrix model with the potential yielding the set of times
in the Toda lattice.

Let us speculate on the appearance of the matrix model in a given
context. To this aim consider the phase space as a two dimensional
noncommutative plane. Evidently the Planck constant $h$ fixes
the scale of noncommutativity due to the canonical commutation relation
\beq
[x_1,x_2]=ih
\eeq
The Moyal multiplication law for the functions on the phase space
reads
\beq
f(x)\ast g(x)= exp (ih \partial _{x} \partial _{y})_{x=y}g(y)f(x)
\eeq
Now let us remind the notion of the Morita equivalence. Qualitatively
it means that for the rational noncommutativity ($h=p/q$)
the noncommutative theory is equivalent to the commutative theory
on the rescaled noncommutative manifold. The functions
on the commutative manifold become matrix valued with the size
of the matrix fixed by the noncommutativity.

Let us apply these circle of ideas to the generic dynamical system.
Suppose that the Planck constant is a rational number $h=\frac{1}{N}$.
If N is large we are dealing with the quasiclassical approximation.
Applying the Morita equivalence, all functions become
$N \times N$ matrix valued and the system becomes effectively classical.
Therefore Morita transform maps the quantum system into
the classical matrix theory. The "classical" partition function
of the matrix model is the prepotential for the dynamical system.

One more indication of the effective classicality of the matrix model
follows from the matrix relation $[M^{+},M]=0$ which
corresponds to the "classical" commutation relation. On the other hand
the corresponding Planck constant in the Morita equivalent system
is $\tilde{h}=N$ which indicates the  regime
of the strong noncommutativity on the phase plane. Along this
viewpoint one could expect the appearance of the peculiar
objects relevant for the such regime, namely the noncommutative
solitons \cite{nsol} which in the simplest case
correspond to    the projector operators $|n><n|$.
One the other hand we have already remarked that the
coherent states also have the description in terms of the
projector operators on the homogeneous spaces. This implies
the intriguing conjecture to consider the matrix model as
the system of the interacting noncommutative solitons.

Let us make one more remark concerning the different interpretations of the
prepotential. It has a interpretation of the energy in the matrix model,
action for the c=1 string and somewhat involved interpretation
in the context of particle dynamics.  This could  presumable have the following
picture behind; it is energy for the membrane and therefore
the action for string. With this picture in mind
the matrix model corresponds to the M(atrix) model representation
for the membrane. This is in a rough agreement with the chain of
T duality transformations of the phase space. Let us also note
that  in the context of c=1 string "times" correspond
to the vacuum expectation values of tachyons with the different
momenta which could be related to the known phenomena of an
appearance of the noncommutative solitons(coherent states) as the result of
the tachyon condensation \cite{sen}.

6. To conclude we have argued that the  Toda lattice
governs both the deformations of the
complex surfaces and analytic curves.
These deformations can be considered as the motion
in the space of the complex and real dynamical systems
respectively or as a kind of RG flows in a peculiar system. The dynamics
with respect to the time $t_0$ can be considered as the
RG evolution with $t_0$ playing the role of the scale factor.
One could also expect that the discrete symmetries
could provide an important tool for investigation
of the RG flows like in \cite{discrete}.

The Baker-Akhiezer function was definitely identified
as the coherent wave function in the action-angle representation
and the generating functional has the interpretation
of the generating function of the canonical transformation from
holomorphic to action-angle representation. The role of the quantization
of the initial dynamical system
in the integrability framework is clarified.

The author is grateful to V. Roubtsov, P. Wiegmann and A. Zabrodin for
the interesting discussions and University of Angers
where the paper has been completed for the hospitality.
The work was partially
supported by grants INTAS-99-1705 and CRDF-RP1- 2108
and by grant for senior scientist fellowship of CNRS 2000.

\end{document}